\documentclass{IEEEcsmag}

\usepackage[colorlinks,urlcolor=blue,linkcolor=blue,citecolor=blue]{hyperref}
\expandafter\def\expandafter\UrlBreaks\expandafter{\UrlBreaks\do\/\do\*\do\-\do\~\do\'\do\"\do\-}
\usepackage{upmath,color}

\jvol{XX}
\jnum{XX}
\paper{8}

\begin{document}


\title{Towards Intelligent Systems for Battery Management: A Five-Tier Digital Twin Architecture}

\author{Tianwen Zhu, Hao Wang, Zhiwei Cao}

\affil{Nanyang Technological University, Singapore, 639798
}

\author{Simon See}

\affil{NVIDIA AI Tech Centre, Singapore, 199589
}

\author{Yonggang Wen}

\affil{Nanyang Technological University, Singapore, 639798}





\markboth{THEME/FEATURE/DEPARTMENT}{THEME/FEATURE/DEPARTMENT}

\begin{abstract}\looseness-1
As digital twin technologies are increasingly incorporated into battery management systems to meet the growing need for transparent and lifecycle-aware operation, existing battery digital twins still suffer from fragmented operational processes and lack an architectural perspective to coordinate modeling, inference, and decision-making throughout the battery lifecycle. 
To this end, we develop a unified five-tier battery digital twin framework that integrates key functionalities into a coherent pipeline and facilitates a clearer architectural understanding of digital twins. The five-tier comprises geometric modeling, descriptive analytics, physics-informed prediction, prescriptive optimization, and autonomous control. 
In quantitative evaluation, the resulting architecture achieves high-fidelity multi-physics calibration with 0.92\% voltage and 0.18\% temperature prediction error, and provides state-of-health estimation with 1.09\% MAPE and calibrated uncertainty.
As the first battery digital twin system empowered by the NVIDIA ecosystem with physics-AI technologies, our proposed five-tier framework shifts battery management from reactive protection to an interpretable, predictive, and autonomous paradigm, paving the path to develop next-generation battery management and energy management systems. 
\end{abstract}

\maketitle

\chapteri{B}attery Management Systems (BMSs) serve a critical function in ensuring the safety and performance of electric vehicles and energy storage systems.
With the increasing scale of battery pack integration and the growing complexity of operating conditions, BMSs should not only ensure basic safety protection but also progressively expand their capabilities to meet the pressing demand for higher-level intelligent management.
The indispensable role of BMSs is driven by the rapid expansion of energy storage applications, expected to reach around 442 GWh globally by 2030 \cite{BNEF2023storage}.
This accelerating deployment is increasing not only capacity demands but also introducing system complexity.
For instance, EV battery packs contain thousands of cells operating under fluctuating loads, while grid-scale systems integrate heterogeneous batteries with diverse degradation patterns \cite{cheng2016consumer}. These complexities demand BMSs capable of coordinating cell behavior and maintaining resilience under varying demand and supply conditions.

In practice, conventional BMSs remain limited in their sensing and prediction capabilities, primarily due to their reliance on a narrow range of sensor-accessible parameters.
First, perception is limited because typical BMS architectures based on embedded PLCs and ECUs can process only a narrow set of measurable signals \cite{muneer2025way}, providing limited visibility into latent states like State of Charge (SOC), or State of Health (SOH) \cite{cui2025towards}. 
This hardware limitation leads most existing BMSs to adopt rule-based designs with threshold-triggered protective actions for short-term safety.
Second, predictive capability is limited because present BMSs mainly rely on simplified physics models that capture only short-term electrical responses while neglecting electrochemical degradation mechanisms. Consequently, they cannot simulate state trajectories under varying load and thermal conditions, making long-term health evolution prediction unattainable. 


To overcome these limitations, researchers have introduced the concept of digital twins into BMSs, integrating multi-source data, multi-physics modeling, and advanced AI methods to establish a dynamic mapping between the physical and virtual domains. A digital twin is a synchronized digital replica of the physical system that combines the asset itself, a high-fidelity virtual model, and bi-directional data flows to maintain alignment \cite{ali2021cognitive}. 
In the battery domain, digital twins typically leverage physics-based electrochemical models and multi-node thermal models to capture spatiotemporal distributions of temperature and lithium concentration.
Because such high-fidelity simulations are computationally intensive, they are often executed on cloud or high-performance computing platforms, with the results synchronized back to the edge BMS for real-time decision support \cite{xie2024dual}.
As a result, current solutions tend to cover one or a few functional layers in isolation separately.


Despite recent progress, the fundamental gap in current battery digital twin frameworks is the absence of a unified, multi-tier architecture that organizes fragmented capabilities and workflows into a coherent whole.
Building on this overarching architectural void, three specific gaps emerge.
First, lacking synergy between data-driven models and physics-based models makes it difficult to simultaneously achieve high predictive accuracy and physical interpretability. In practice, many battery digital twin implementations rely on black-box data-driven approaches that capture correlations but ignore physical laws, while physics-based electrochemical or thermal models offer mechanistic insight but remain computationally intensive and struggle to represent degradation behaviors under practical cycling conditions.
Second, most existing battery digital twins lack self-evolving capability. They still depend heavily on offline analysis, which prevents timely updates of the virtual model. As a result, the digital twin cannot continuously adapt to real-time operating data, leading to gradual divergence from the physical battery under dynamic conditions. 
Third, current systems lack decision validation capabilities. Although many digital twins can generate accurate predictions of battery behavior, they seldom verify how these predictions translate into effective operational decisions. Thus, the digital twin cannot ensure that its recommendations, such as charging or scheduling strategies, consistently improve real-world performance.


To bridge these gaps, we propose a highly intelligent five-tier battery digital twin architecture empowered by the NVIDIA ecosystem, systematically organizing the full spectrum of digital twin capabilities.
This battery digital twin integrates real-time data assimilation at the descriptive tier, multi-physics simulation for accurate forecasting at the predictive tier, optimization algorithms for prescriptive control, and autonomous closed-loop operation at the highest tier, forming a continuous pipeline from monitoring to intelligent actuation. 
Electrochemical and thermal dynamics are captured in physics-based simulators and synchronized within NVIDIA Omniverse, with NVIDIA SimReady assets streamlining standardized model construction, while NVIDIA PhysicsNeMo provides the physics-informed learning engine that accelerates simulation and forecasting.
Unlike conventional designs that separate monitoring, forecasting, and control into isolated modules, our digital twin establishes an evolving digital intelligence across the entire battery lifecycle, enabling continuous synchronization and self-evolution between the physical and virtual systems, thereby delivering enhanced safety and reliability for next-generation EVs and energy storage applications.

The contributions of our approach are as follows.
\begin{itemize}
    \item To the best of our knowledge, this is the first article that systematically proposes and introduces a unified five-tier digital twin system for intelligent battery management. Each tier is clearly defined to address the gaps in existing BMS approaches.

    \item The system uniquely employs calibrated multi-physics modeling with PIML to achieve interpretable tracking of internal battery states and latent variables, which directly enhances the physical interpretability of battery management and supports proactive decision-making.  

    \item We present the first battery digital twin system implemented within the NVIDIA ecosystem, leveraging Omniverse, SimReady assets, and PhysicsNeMo for end-to-end geometric modeling, multi-physics simulation, and physics-informed learning. Our implementation delivers high simulation fidelity, achieving low voltage and temperature prediction error, and produces robust SOH estimates with quantified uncertainty.
\end{itemize}

\vspace{-0.3cm}
\section{Preliminaries}
In this section, we first review the common BMS architectures. We then introduce multi-physics modeling approaches for electrochemical, thermal, and degradation processes. Finally, we summarize key simulation platforms and validation methods that bridge virtual models with real-world operation.

\subsection{Overview of BMS}
BMSs are designed to monitor and protect batteries. In practice, a BMS integrates subsystems for signal acquisition, central control, circuit protection, and communication with upper-level energy management systems \cite{gabbar2021review}. Despite these capabilities, conventional implementations remain limited in intelligence, as they rely heavily on rule-based logic where protective actions such as disconnection, cooling, or cell balancing are triggered only after predefined safety thresholds are exceeded \cite{fioravanti2020predictive}. 
While such designs mitigate immediate risks, they remain inherently reactive and inefficient, motivating the need for more advanced and predictive BMS solutions.

\subsection{Battery Multi-physics Modeling Techniques}

Battery physics modeling sets up the foundation for estimating and predicting battery dynamic behavior based on electrochemical, chemical and mechanical principles.
At the electrochemical level, broadly used models include the Single Particle Model (SPM) and the Doyle–Fuller–Newman (DFN) model \cite{doyle1993modeling}.
The SPM simplifies battery dynamics by considering single representative particles, offering computational efficiency, while the DFN model provides detailed insights by simulating complex lithium-ion transport and electrochemical reactions across battery electrodes.

Thermal modeling is a key component that characterizes heat generation, transfer, and dissipation during battery operation, typically formulated from energy conservation laws. Lumped thermal models \cite{alkhedher2024electrochemical} are widely adopted due to their simplicity and computational efficiency in estimating temperature dynamics, thereby supporting effective thermal management strategies essential for safety and reliability. Battery degradation modeling, on the other hand, focuses on long-term aging driven by mechanisms such as solid electrolyte interphase growth, lithium plating, and particle cracking. Advanced modeling methods, such as those introduced by Wang et al. \cite{wang2018review}, explicitly couple these processes to capture the complex interactions that govern performance decay and material loss. These integrated multi-physics modeling plays an important role in accurately predicting battery lifetime and performance, emphasizing the importance and challenge of incorporating comprehensive multi-physics models within digital twin architectures. It is worth mentioning that both the thermal models and degradation models can be coupled within the multi-physics model by introducing new source functions.

\vspace{-0.3cm}
\subsection{Simulation Platforms and Validation Methods}
High-fidelity simulation platforms are a trustworthy way to validate battery digital twins, ensuring accurate representation of electrochemical, thermal, and mechanical behaviors under diverse conditions. Besides well-known tools like PyBaMM and COMSOL, other notable platforms include ANSYS Fluent for computational fluid dynamics and thermal analysis, and MATLAB/Simulink for dynamic system modeling and control strategy simulations. These platforms support analyzing and simulating battery operation, facilitating comprehensive virtual testing without the necessity for costly physical prototype validations.

Hardware-in-the-Loop (HIL) and Software-in-the-Loop (SIL) are two main complementary simulation paradigms that establish a bridge between virtual models and real-world systems \cite{bui2019advanced}. In SIL, controller algorithms and software modules are embedded within a virtual simulation environment, allowing rapid prototyping, algorithm debugging, and iterative design without the need for physical hardware. Moreover, HIL extends this principle by incorporating real physical components into the simulation loop, thereby exposing algorithms to realistic operating conditions and hardware constraints. By combining these two approaches, digital twins benefit from a continuous validation pipeline that spans from early-stage software verification to hardware-level performance testing, laying the foundation for reliable predictive maintenance and operational optimization of batteries.

\section{System Design of the Digital Twin}

In this section, we present the intelligent five-tier digital twin system for battery management. First, we introduce the overall system architecture. Then, we articulate several potential applications of this digital twin in predictive and prescriptive battery analysis.

\subsection{Architecture Overview}

\begin{figure}[tbp!]
  \centering
  \includegraphics[width=0.48\textwidth]{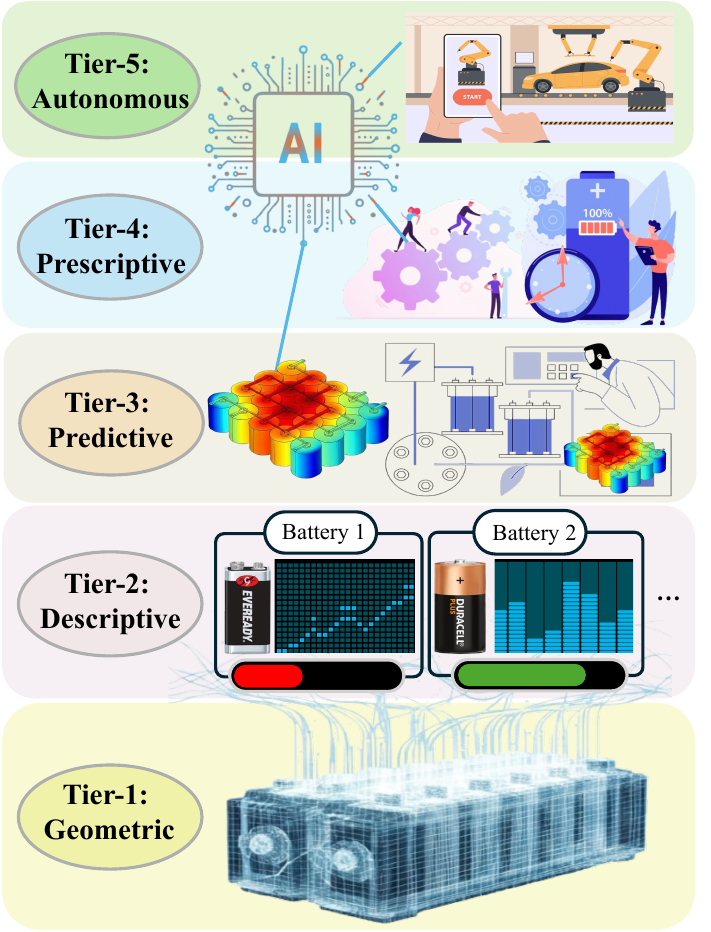} 
  \vspace{-0.1cm}
  \caption{Overview of the proposed five-tier digital twin intelligence system, integrating geometric modeling, descriptive analytics, predictive forecasting, prescriptive optimization, and autonomous control for intelligent battery management.} 
  \vspace{-0.2cm}
  \label{high level framework}
\end{figure}

The architecture of our intelligent five-tier digital twin system is shown in \autoref{high level framework}. Each layer is constructed upon the foundation of the previous one within a unified intelligent system.
From bottom to top, Tier~1 (Geometric) builds 3D representations of battery cells, modules, and packs, implemented in NVIDIA Omniverse as the spatial backbone, supported by NVIDIA SimReady assets that provide standardized, simulation-ready battery components; Tier~2 (Descriptive) binds real‑time sensor streams to this geometry to create a live, data‑rich twin that visualizes temperature distributions, voltage/current profiles, and operating conditions; Tier~3 (Predictive) leverages PIML to forecast capacity‑degradation trajectories, estimate SOH and RUL, and quantify thermal‑runaway risk under diverse scenarios; Tier~4 (Prescriptive) translates these forecasts into optimal operating strategies, such as fast‑charging protocols, cell‑balancing schedules, and cooling‑system set‑points, while enforcing safety and operational constraints; and Tier~5 (Autonomous) enables the system to achieve closed-loop, AI-driven control, where decisions are executed independently and models adapt through continual learning.
NVIDIA Omniverse underpins Tier 1 and Tier 2 as the shared spatial and data plane, augmented by NVIDIA SimReady standardized battery assets, while PhysicsNeMo powers Tier 3 to Tier 5 for learning, prescription, and autonomy.

This intelligent system is enabled by three core modules spanning across tiers, with the overall organization illustrated in \autoref{fig:DTframework}: (i) a virtual 3D environment powered by NVIDIA Omniverse and NVIDIA SimReady for geometric modeling and visualization, (ii) a multi-physics simulation engine for high-fidelity battery state estimation and prediction, and (iii) an AI engine leveraging NVIDIA PhysicsNeMo with PIML to tighten synchronization between the physical battery and its digital replica. 

\subsubsection{Virtual 3D Environment Module}
The virtual 3D environment underpins Tier 1 (Geometric) and Tier 2 (Descriptive) with NVIDIA Omniverse serving as the foundational platform for constructing high-fidelity virtual replicas of battery systems.
As illustrated in \autoref{fig:DTframework}, this module forms the core of the Visualization Layer, where USD files enable standardized geometric modeling spanning from cell components to complete battery system assemblies with many packs. 
Within this system, this module is designed based on a USD-based 3D scene environment, which is completely programmable using a Python script to add or remove 3D objects and determine their spatial location. In our settings, the scene graph captures the full battery hierarchy from cell to module to pack to represent a 20 kWh-level lithium-ion battery energy storage system.
To accelerate and standardize the construction of such virtual battery assemblies, we leverage NVIDIA SimReady assets, which provide preconfigured, simulation-ready models of cells, modules, and packs that integrate natively with the USD-based Omniverse environment.

On top of this geometric structure of an energy storage system, the module adds a semantic schema that assigns meaning to each element. While the scene graph defines how cells, modules, and packs of the batteries are arranged in space, the semantic schema describes what each object represents and how it should be interpreted. For instance, a node can be marked as a temperature sensor, linked to the quantity it measures, and associated with a unit such as degrees Celsius. In this way, the schema ensures that the 3D scene is not just a geometric model but also a machine-readable map of physical roles and properties.
To ensure that these semantic descriptions are usable across different scenarios in the digital twin, the module also fixes global coordinate frames, standardizes unit systems, records sensor locations, and assigns unique identifiers. These definitions make it possible to map live telemetry onto the 3D geometry without ambiguity and to exchange data consistently among different simulation and control tools.

\begin{figure}[tbp!]
  \centering
  \includegraphics[width=0.48\textwidth]{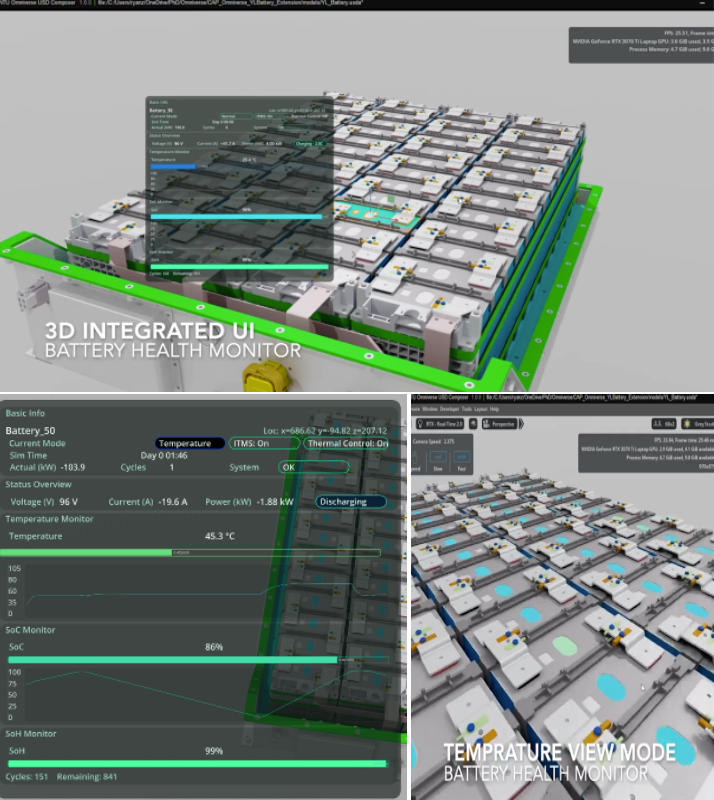} 
  \vspace{-0.1cm}
  \caption{Battery digital twin visualization in NVIDIA Omniverse, showing pack-level monitoring with predictive analytics. The system integrates real-time health indicators with SOH, RUL, and temperature prediction modules to support proactive safety and maintenance decisions.} 
  \vspace{-0.2cm}
  \label{fig:Omni}
\end{figure}

In addition, the 3D environment module defines live data mappings that connect telemetry streams and simulation outputs to the corresponding geometric entities. These mappings specify how data is transmitted and processed, including communication topics, update rates, and data types, while providing standardized input–output interfaces to connect with other tiers. The module also exposes read/write APIs for Tier 3 predictive services to present measured, estimated, and predicted results such as SOH and RUL in the 3D scene, while for Tier 4 controllers to visualize recommended setpoints.
\autoref{fig:Omni} illustrates this integration within the Omniverse digital twin interface. This figure presents a 3D battery pack model augmented with live sensor dashboards and overlays of predicted SOH, RUL, and temperature variations results.
Through this design, the shared 3D scene serves as the central hub where all information about the operational condition, real-time system information, and optimized control commands are integrated, ensuring consistency and coordination inside the battery digital twin.

\subsubsection{Multi-physics Simulation Engine}
The multi-physics simulation engine underpins Tier 3 (Predictive) by enabling high-fidelity forecasting of battery dynamics through coupled electrochemical, thermal, and mechanical modeling.
As shown in \autoref{fig:DTframework}, this module constitutes the core of the Model Layer and Simulation Engine, which serve as the analytical backbone of the overall system.
Simulation outputs are streamed to Omniverse for live visualization and are used as supervised and physics constrained signals for model training and inference in PhysicsNeMo.

The simulation engine integrates these physics domains into a unified, co-evolving environment, ensuring that the digital twin reproduces battery behavior in a physically consistent manner. For example, electrochemical reactions and ohmic resistance heat are identified as different heat generation sources that result in thermal dynamic variations. The rising temperature inside the cell then impacts the reaction kinetics and transport processes as described by the Arrhenius equation \cite{laidler1984development}, reshaping voltage responses and accentuating spatial non-uniformities across cells. 
By resolving these cause-and-effect loops, the multi-physics simulation engine can capture critical battery behaviors such as hot-spot formation, heat spread, and performance shifts under varying loads.

This multi-physics simulation engine not only aligns with experimentally observed behaviours of actual battery systems but also produces high-quality synthetic datasets for the battery digital twin. It also conducts calibrated simulation campaigns in which operating profiles, ambient conditions, thermal management strategies, and aging states are systematically synchronised, generating multi-domain information in time series such as voltage, current, temperature fields, concentration distributions, and derived health indicators. Due to these outputs being computed in a coupled physics model, they are physically interpretable and can be used to deduce other critical latent states that are hard to observe using sensors. This enriches the battery health-related information, making it particularly valuable for supervised learning and benchmarking. Furthermore, the data reliability of this simulation engine is ensured through real-time calibration against experimental measurements to make the simulation responses match well with the observed responses. Additionally, physical consistency checks based on universal physical principles, such as energy conservation laws, are utilised to further reduce the mismatch between the digital twin battery and the physical battery.

The augmentation datasets generated from the multi-physics simulation engine support the formation and enhance the performance of digital twin by (i) providing training supplementary datasets for downstream PIML tasks, (ii) forming operational scenario libraries that reveal trade-offs among various factors, such as charging rate, round-trip efficiency, and thermal effect and (iii) verification of control policies under rare or hazardous conditions before safe transfer to HIL testbeds. By combining these techniques, the simulation engine elevates the twin from a passive mirror of sensor streams into an active experimental platform for intelligent battery management, monitoring and prognostics.

\subsubsection{AI Engine Module}

\begin{figure*}[tbp!]
  \centering
  \includegraphics[width=0.99\textwidth]{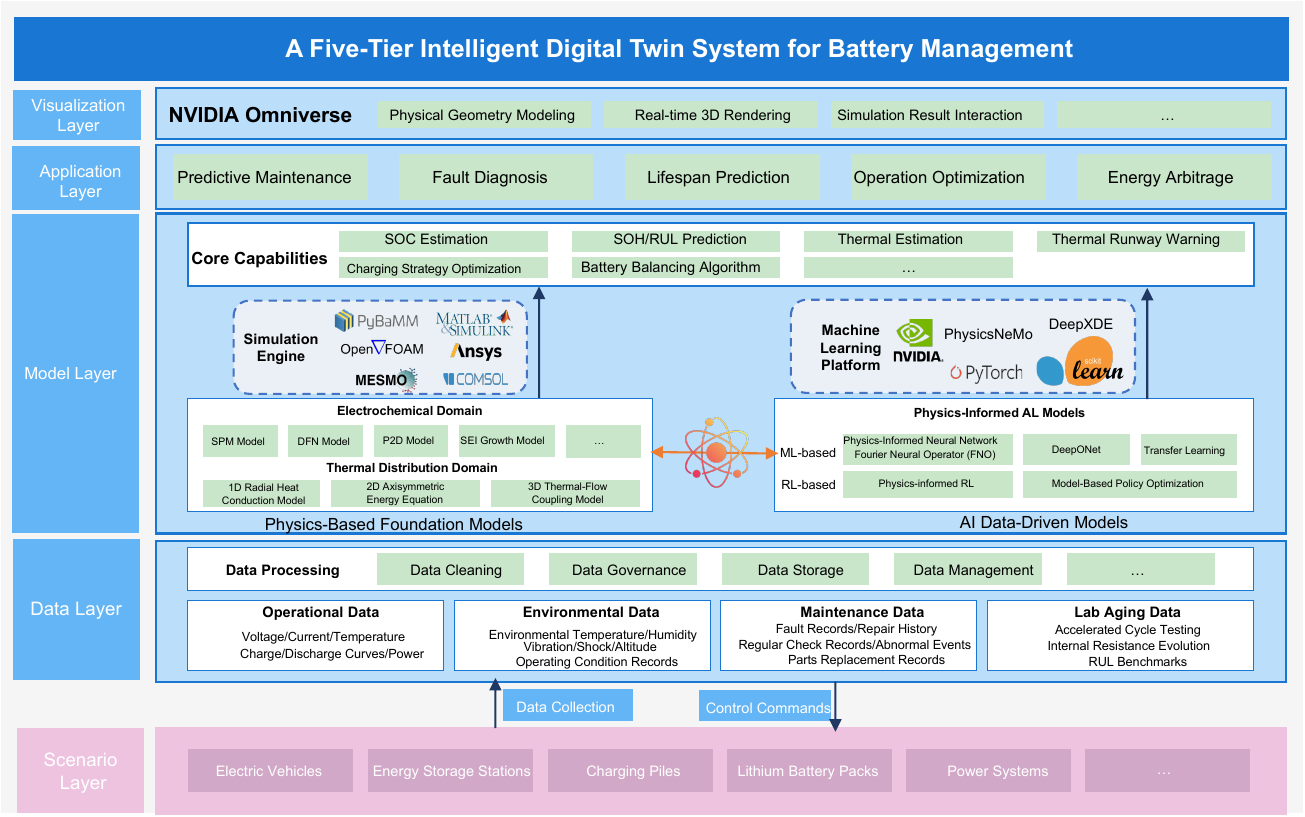} 
  \caption{Technical architecture of the five-tier intelligent digital twin system for battery management. The comprehensive technical architecture includes multi-physics modeling tools, advanced simulation platforms, and physics-informed AI methodologies such as PhysicsNeMo. The layered structure enables systematic data integration, accurate predictive simulations, and effective prescriptive strategies, thus supporting proactive battery health management throughout its lifecycle.} 
  \vspace{-0.2cm}
  \label{fig:DTframework}
\end{figure*}

The AI engine supports Tier 4 (Prescriptive) and Tier 5 (Autonomous) by providing intelligent optimization and adaptive control capabilities for the battery digital twin.
As illustrated in \autoref{fig:DTframework}, it represents the intelligence core of the overall system within the Machine Learning Platform, where advanced AI methods complement physics-based modeling.
In our digital twin system design, NVIDIA PhysicsNeMo serves as the foundation of the AI engine, enabling the battery digital twin to deliver real-time intelligent estimation, prediction and management while preserving physical consistency and interpretability. PhysicsNeMo follows a PIML paradigm that incorporates governing principles such as charge conservation, electrochemical kinetics, and thermodynamic constraints are embedded directly into neural architectures and training objectives, ensuring both estimations and predictions remain physically plausible and trustworthy rather than purely data-driven. PhysicsNeMo implements neural PDE solvers to accelerate electrochemical and thermal dynamics by several orders of magnitude compared to classical finite-element methods. Meanwhile, transformer-based or other sequence-learning models capture long-term dependencies that characterize degradation trajectories. Together, these methods provide robust forecasts of voltage response, temperature evolution, SOH, RUL, and associated uncertainty at battery cell, module, and pack levels.  

Building on its predictive accuracy, the AI engine also supports prescriptive analytics by turning predictions into concrete operating strategies. For example, reinforcement learning with experience replay utilizes past driving or charging information to gradually improve the charging strategy, and thermal management is scheduled as batteries age. Transfer learning accelerates the roll-out of new systems by reusing models trained on similar fleets, reducing the necessity of model retraining. In addition, AI-driven multi-objective optimization methods are applied to generate clear trade-off curves that show, for example, how increasing the charging rate may reduce efficiency or raise cell temperature. This information gives operators and upper-level controllers practical choices, facilitating them to select strategies that well-fit their performance and safety requirements under varying conditions.

From a system-level perspective, the AI engine drives autonomous operation by turning forecasts and optimizations into direct control actions. It utilized continual learning to adapt operational policies as usage conditions vary, and employs fault-tolerant mechanisms that keep the system functional when components fail. Before new decisions are enacted, the engine runs fast predictive simulations to conduct risk detections such as thermal runaway or internal short-circuit. To validate these learned policies work effectively on real battery systems, the AI engine is coupled with HIL platforms (e.g., OPAL-RT), which execute virtual commands against real controllers and components. In this way, optimized policies are validated under realistic constraints before deployment, ensuring that the digital twin can operate as a self-governing battery manager across its lifecycle. 

\subsection{Practical Applications of The Proposed Intelligent System}

The capabilities of our intelligent digital twin system enable transformative applications across battery management, through two primary domains that collectively address intelligent battery operations.

\subsubsection{Predictive Battery Health Management}

The most immediate value of the proposed battery digital twin lies in its ability to transform health prediction into a continuous and adaptive process. Instead of relying on sparse snapshots, SOH and RUL are evaluated in real time, adapting to variations in usage and environment. This approach improves predictive accuracy and produces risk-aware outputs expressed as probability distributions rather than deterministic forecasts, enabling operators to plan maintenance proactively with quantified confidence \cite{gama2022data}. Such predictive intelligence ensures that maintenance actions are both timely and cost-effective, reducing unexpected downtime and improving overall system reliability.

Beyond prediction, the digital twin enables early detection of degradation far in advance of conventional monitoring tools. By identifying subtle indicators of failure, such as cell imbalance or incipient thermal runaway, it allows operators to isolate or rebalance systems before issues escalate. This capability supports practical applications across fleets and grid storage by preventing propagation of failures and guiding warranty or replacement decisions based on emerging degradation patterns. By distinguishing short-term performance drift from irreversible damage, the system ensures that minor fluctuations are managed efficiently while critical risks are addressed without delay, strengthening both safety and operational resilience.

\subsubsection{Battery Repurposing and Value Maximization}

Beyond health prediction, the digital twin enables actionable strategies to maximize battery value across its entire lifecycle. During operation, forecasts are translated into adaptive charging protocols that shorten charging duration while keeping temperature rise and energy loss within safe limits, and intelligent thermal management that dynamically adjusts airflow and coolant flow according to predicted heat loads. These strategies not only improve energy efficiency but also mitigate thermal stress that accelerates degradation. At the system level, the twin orchestrates load distribution across modules with different health states. Instead of uniform current sharing, it redistributes demand away from overstressed or degraded packs, thereby protecting weaker components while extracting maximum capacity from healthier ones. This capability is particularly valuable in grid-scale storage plants integrating batteries of mixed chemistries and vintages, where conventional rule-based control often leads to underutilization of robust modules and premature aging of already stressed ones \cite{koltermann2024improved}.
Finally, by preserving complete degradation histories, the digital twin provides accurate assessments of residual capacity at end-of-life. This enables reliable decisions for second-life deployment in stationary storage or recycling, ensuring safe, efficient, and sustainable reuse while maximizing residual value.


\section{System Implementation and Evaluation}

In this section, we implement the proposed five-tier intelligent digital twin system based on NVIDIA ecosystem, focusing on model calibration and health prediction for practical battery management.

\subsection{Problem Settings}

XJTU battery dataset \cite{wang2024physics} is used to demonstrate practical battery management. The dataset comprises experiments on 55 lithium-ion batteries ($\mathrm{LiNi_{0.5}Co_{0.2}Mn_{0.3}O_{2}}$), each with a nominal capacity of $2000\,\mathrm{mAh}$ and a nominal voltage of $3.6\,\mathrm{V}$. The experiments cover six charge and discharge regimes that emulate realistic operating conditions, including constant C rates, variable-rate profiles, randomized usage, and a satellite-use pattern.

The problem setting is to implement two key functions of the proposed system. 
The first task is to calibrate a coupled electrochemical–thermal model that can accurately simulate voltage and temperature dynamics under diverse cycling conditions. 
The second task is to develop predictive models for battery SOH estimation with quantified uncertainty, enabling risk-informed decision support for lifecycle management.

\subsection{Realization of the Five-Tier System}
We instantiate these tasks by mapping them to Tier 1–5 of the proposed system. Figure~\ref{fig:Omni} illustrates how Tiers~1 and 2 are realized in the Omniverse-based 3D environment, where geometric models are bound with telemetry and simulation data to visualize battery states and predictions. 
We additionally leverage NVIDIA SimReady assets to accelerate the construction of standardized, simulation-ready battery components for Tiers~1 and 2, ensuring consistent geometric and physical representations within the Omniverse environment. 
Building on this foundation, Tier~3 is enabled through multi-physics modeling and calibration, while Tiers~4 and 5 leverage AI engines to provide predictive and prescriptive intelligence with uncertainty quantification. 

\subsubsection{Tier 3: Multi-physics Modeling and Calibration}
We integrate the SPM and DFN electrochemical models with a lumped thermal model using PyBaMM as the modeling platform \cite{sulzer2021python}. The SPM offers computationally efficient simulations suitable for rapid, real-time predictions, while the DFN model provides detailed representation of lithium-ion transport processes and reaction kinetics. A lumped thermal model is coupled with the DFN to accurately capture thermal dynamics under diverse operating conditions.

To ensure fidelity with respect to experimental battery performance, we apply Bayesian optimization for parameter calibration. This procedure adjusts over 15 model parameters (e.g., diffusivities, reaction rates, conductivities, electrode geometries, thermal conductivities) by minimizing discrepancies between simulated and measured voltage and temperature. Through iterative refinement, the calibrated models achieve strong consistency with experimental observations, forming the predictive backbone of the digital twin.

\subsubsection{Tier 4: Physics-informed Neural Network}
Leveraging NVIDIA PhysicsNeMo, we construct PINNs trained on data generated from the calibrated multi-physics simulations. The loss function integrates both data-driven accuracy terms and physics-based constraints, penalizing violations of electrochemical principles and thermal equilibrium. By enforcing these constraints via automatic differentiation at strategically selected points, the PINN captures complex battery dynamics while maintaining physical consistency. This enables Tier~4 prescriptive functions by providing physically trustworthy predictions for SOH and RUL.

\subsubsection{Tier 5: Uncertainty Aware Decision-Making}
To support automatically decision-making in Tier~5, we incorporate Deep Autoencoding Gaussian Mixture Model (DAGMM) for uncertainty quantification. DAGMM combines an autoencoder with a Gaussian Mixture Model (GMM) to evaluate the likelihood of latent representations of operational data. The resulting energy-based score provides a principled uncertainty metric, identifying anomalies and quantifying confidence in predictions. By capturing uncertainty alongside predictions, this implementation equips the digital twin with the ability to support risk-informed management decisions.

\subsection{Evaluation Results}
Rigorous evaluations are given to demonstrate the superior predictive accuracy and reliability of the proposed digital twin system through multi-physics model calibration and uncertainty quantifications.

\begin{figure}[tbp!]
  \centering
  \begin{subfigure}{0.23\textwidth}
    \includegraphics[width=\textwidth]{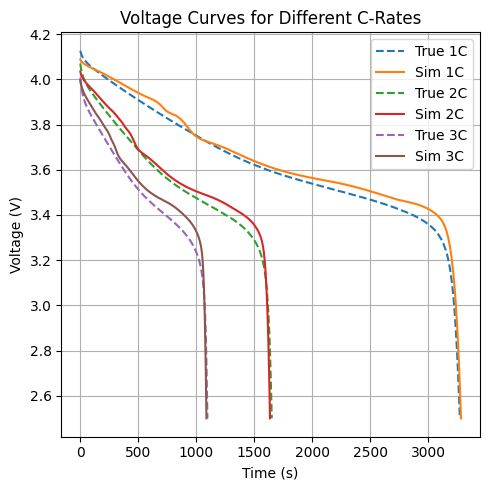}
    \caption{Voltage prediction under 1C, 2C, and 3C discharge rates.}
    \label{calibration_a}
  \end{subfigure}
  \hfill
  \begin{subfigure}{0.23\textwidth}
    \includegraphics[width=\textwidth]{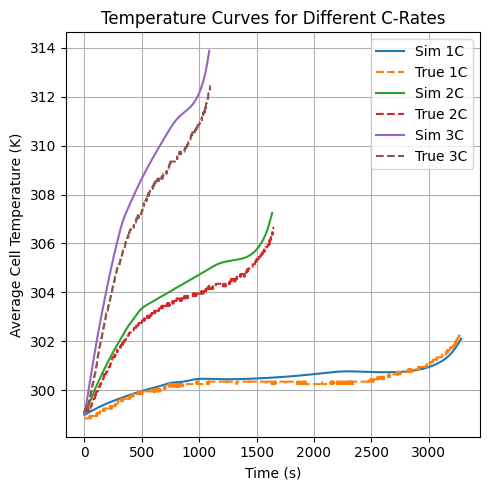}
    \caption{Temperature prediction under 1C, 2C, and 3C discharge rates.}
    \label{calibration_b}
  \end{subfigure}
  \caption{Multi-physics calibration results against the XJTU battery dataset after Bayesian optimization.}
  \label{calibration}
\end{figure}

For the multi-physics calibration, our model demonstrated good estimation accuracy compared with the XJTU battery dataset as shown in \autoref{calibration}. Across ten repeated trials, the calibrated model achieved an average voltage prediction MAPE of $0.92 \pm 0.15\%$ (1C), $1.06 \pm 0.17\%$ (2C), and $1.57 \pm 0.21\%$ (3C). For temperature prediction, the MAPE reached $0.07 \pm 0.01\%$ (1C), $0.18 \pm 0.01\%$ (2C), and $0.39 \pm 0.05\%$ (3C). These results validate the effectiveness of the multi-physics model in accurately reproducing voltage and temperature behaviors of batteries.

\begin{figure*}[t]
    \centering
    \includegraphics[width=\textwidth]{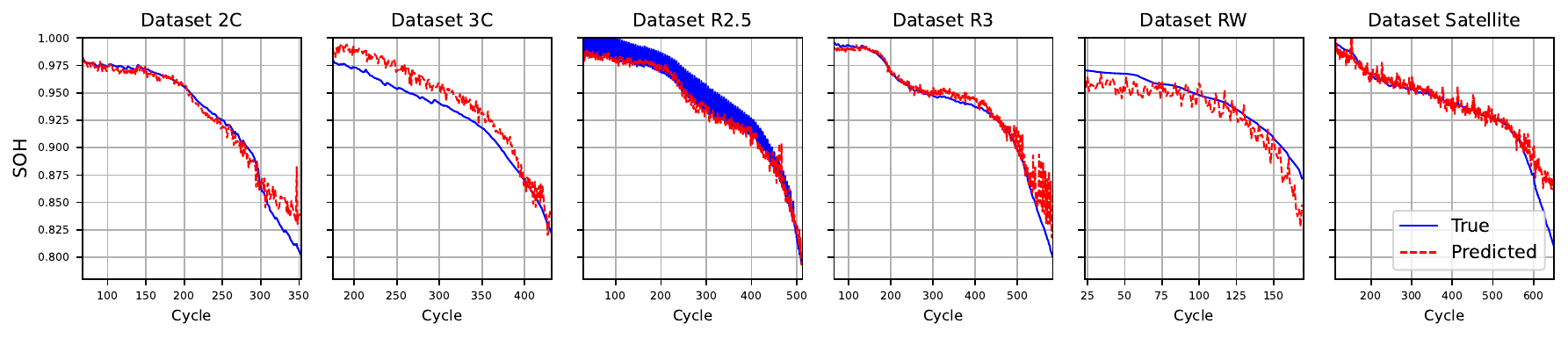}
    \caption{Prediction results across six charging protocol datasets. Each panel plots battery SOH versus cycle index for a distinct dataset (2C, 3C, R2.5, R3, RW, Satellite). The red curve shows the true SOH labels and the blue curve shows predictions from our model. Panels share comparable axis limits to enable fair visual comparison across datasets. The results indicate that predictions closely follow the true trajectories under both constant rate and randomized profiles.}
    \label{fig:true-predict}
\end{figure*}

\begin{figure}[tbp!]
  \centering
  \includegraphics[width=0.45\textwidth]{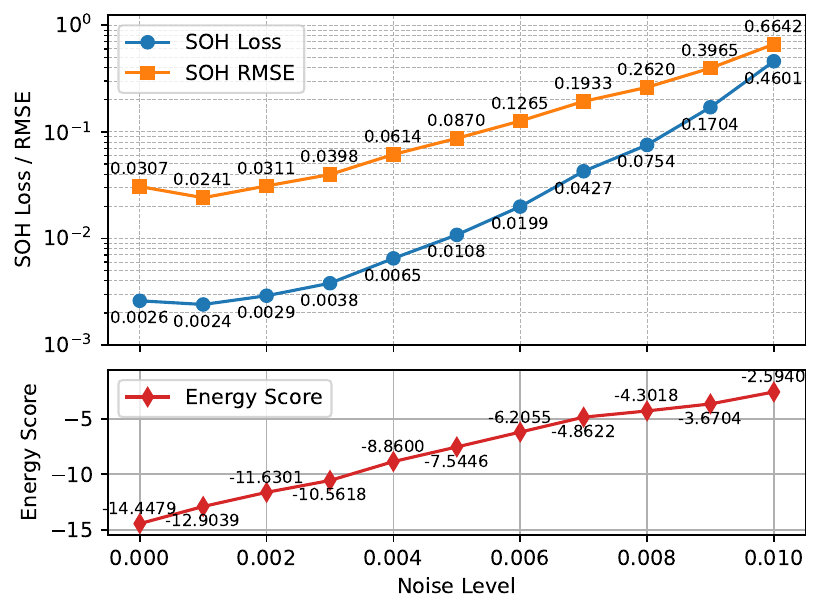} 
  \vspace{-0.1cm}
  \caption{Impact of noise injection on prediction error and the proxy metric for uncertainty.} 
  \vspace{-0.2cm}
  \label{dagmm}
\end{figure}

Building upon the calibrated simulations, we can further assess the predictive performance of our PINN-based SOH prediction model, including its capability to quantify prediction uncertainty. As shown in \autoref{fig:true-predict}, our model demonstrated robust predictive accuracy in SOH, across ten runs achieving MAPE at $1.09 \pm 0.04 \%$. Most importantly, the integrated uncertainty quantification via the DAGMM provided reliable indications of prediction confidence. As shown in \autoref{dagmm}, the energy-based uncertainty scores exhibited a strong positive correlation with the actual prediction errors, effectively highlighting scenarios where predictions were less reliable due to data distribution shifts.

\section{Future Directions}
In this section, we discuss several future directions that can be explored to further enhance the intelligent battery management using this digital twin architecture.

\textbf{Foundation Models for Battery Intelligence:}
Large Language Models (LLMs) are foundation models trained on massive text corpora using transformer architectures, enabling diverse tasks such as reasoning, summarization, and question answering without task-specific supervision. Beyond language, they serve as general-purpose engines for knowledge representation and code generation across scientific domains. Building on these capabilities, LLMs adapted on battery-related literature present promising opportunities for digital twins through automated knowledge extraction and model generation \cite{zuo2025large}. Future implementations envision specialized battery foundation models that automatically design PINN architectures for specific chemistries, synthesize insights from vast research to identify degradation mechanisms, recommend experimental protocols, and support conversational interfaces for intuitive querying of battery states and explanations of complex degradation phenomena.

\textbf{Blockchain-based Battery Passport Systems:}
The implementation of blockchain in battery lifecycle management offers transformative potential for future battery passport systems, enabling complete traceability and transparency. Distributed ledger technologies will record manufacturing data, operational history, maintenance, and performance metrics as immutable entries, ensuring reliable information sharing across manufacturers, operators, and recyclers \cite{dasgupta2025intelligent}. Smart contracts could automate management decisions such as maintenance scheduling and end-of-life processing, reducing fraud in condition reporting. Blockchain-based passports will further support circular economy initiatives by providing verified health data for second-life applications, allowing batteries retired from vehicles to be reused in stationary storage with confidence. 

\textbf{Differentiable Simulation for Optimal Control:}
The development of fully differentiable battery simulation frameworks enables gradient-based optimization of battery operational strategies directly through physics-based models, eliminating the need for computationally expensive reinforcement learning approaches. Future implementations could leverage automatic differentiation through simulation platforms to enable direct optimization of charging protocols, thermal management strategies, and load balancing decisions with respect to battery health and performance objectives. This approach provides enhanced transparency compared to black-box optimization methods while enabling principled handling of operational constraints through physics-based penalty terms in the optimization objective.

\section{Conclusion}
To promote the transition of BMS towards intelligent systems for achieving full-lifecycle efficient management across diverse energy storage applications, we presented an intelligent battery management concept based on a five-tier digital twin system for autonomous operations. We first highlighted current BMS challenges and the need for advanced digital twin approaches. Then, we introduced the system design, progressing from geometric modeling to autonomous operation, supported by PIML and uncertainty quantification. Its applications span predictive health management, operational optimization, and lifecycle control. Finally, a system implementation validated our design, showing sub-1\% voltage/temperature errors and robust SOH predictions with MAPE below 3\%, demonstrating the system’s potential for interpretable modeling and reliable decision-making.

\bibliographystyle{IEEEtran}
\bibliography{Ref}









\end{document}